\documentclass[prb,aps,twocolumn,floatfix,amsmath,amssymb,superscriptaddress,tightenlines]{revtex4-1}

\usepackage{graphicx}
\usepackage{savesym}
\usepackage{amsfonts}
\usepackage{bm}
\usepackage{color}
\usepackage{hyperref}
\usepackage{subfigure}
\usepackage{wasysym}

\begin{document}

\title{Quantum Kagome Ice}

\author{Juan Carrasquilla}
\email{jcarrasquilla@perimeterinstitute.ca} 
\affiliation{Perimeter Institute for Theoretical Physics, Waterloo, Ontario, N2L 2Y5, Canada} 

\author{Zhihao Hao} 
\affiliation{Department of Physics and Astronomy, University of Waterloo, Ontario, N2L 3G1, Canada} 

\author{Roger G. Melko} 
\affiliation{Perimeter Institute for Theoretical Physics, Waterloo, Ontario, N2L 2Y5, Canada} 
\affiliation{Department of Physics and Astronomy, University of Waterloo, Ontario, N2L 3G1, Canada}

\date{\today} 

\begin{abstract}
\bf{
Actively shought since the turn of the century, two-dimensional quantum spin liquids (QSLs) are exotic phases of matter
where magnetic moments remain disordered even at extremely low temperatures.
Despite ongoing searches, 
QSLs remain elusive, due to a lack of concrete knowledge of the microscopic mechanisms that inhibit magnetic order in real materials.
Here, we study a theoretical model for a broad class of frustrated magnetic rare-earth pyrochlore materials called
``quantum spin ices''.  When subject to an external magnetic field along the [111] crystallographic direction,
the resulting spin interactions contain a mix of geometric frustration and quantum fluctuations in decoupled two-dimensional kagome planes.
Using large-scale quantum Monte Carlo simulations, we 
identify a simple set of interactions sufficient to promote a groundstate
with no magnetic long-range order, and a gap to excitations, consistent with a $Z_2$ spin liquid phase. 
This suggests a systematic experimental procedure to search for two-dimensional QSLs within the broader class of 
three-dimensional pyrochlore quantum spin ice materials.
 }
\end{abstract}

\maketitle

\section*{\uppercase{Introduction}}

In a two-dimensional (2D) quantum spin liquid (QSL) state, strong quantum fluctuations prevent the ordering of magnetic spins, even at zero temperature.
The resulting disordered phase can potentially be a remarkable state of matter, supporting a range of exotic quantum phenomena.  
Some, such as 
emergent gauge structures and fractional charges, are implicated in a wide range of future technologies like high-temperature superconductivity\cite{ANDERSON1987,Lee2006} 
and topological quantum computing.\cite{Ioffe2002}
It is therefore remarkable that, despite extensive examination of the basic theoretical ingredients required to promote
a 2D QSL in microscopic models,\cite{Balents2010,Yan2011}
the state remains elusive, with only a few experimental candidates existing today.\cite{Pratt2011,Han2012} 

Recently, the search for QSL states has turned to consider
quantum fluctuations in the so-called {\it spin ice} compounds.\cite{Gingras_McClarty2013} In these systems, magnetic ions reside on a pyrochlore  
lattice---a non-Bravais lattice consisting of corner-sharing tetrahedra. 
Classical magnetic moments (described by Ising spins) on the pyrochlore lattice can be geometrically frustrated at low temperatures, 
leading to spin configurations that obey the so-called ``ice rules'', 
a mapping to the proton-disorder problem in water ice\cite{Pauling1935}. 
The ice rules result in a large set of degenerate ground states -- a {\it classical} spin liquid with a finite thermodynamic entropy per spin.\cite{Bramwell16112001,Gingras2011}
Two canonical materials, Ho$_2$Ti$_2$O$_7$ and Dy$_2$Ti$_2$O$_7$, have been demonstrated to 
manifest spin ice behaviour,
and experiments and theory enjoy a healthy dialog due to the existence of classical microscopic models
capable of describing a wide range of experimental phenomena.\cite{Bramwell16112001}

Classical spin ice pyrochlores
are conjectured to lead to QSLs in the presence of the inevitable quantum 
fluctuations at low temperatures.\cite{Gingras_McClarty2013,Balents2010}
The effects of certain types of quantum fluctuations on the spin ice state have been investigated 
theoretically\cite{Hermele2004} and numerically,\cite{Argha_Baek_2008,Shannon2012} where they have been demonstrated 
to lift the classical degeneracy 
and promote a three-dimensional (3D) QSL phase with low-energy gapless excitations that behave like photons.\cite{Hermele2004,Argha_Baek_2008}
In several related pyrochlore compounds, particularily Tb$_2$Ti$_2$O$_7$,  Yb$_2$Ti$_2$O$_7$, Pr$_2$Zr$_2$O$_7$, and Pr$_2$Sn$_2$O$_7$,
quantum effects have been observed,
which make them natural candidates to search for such 3D QSLs.\cite{Canals2007,Kimura2013,Fennell2014}
In an attempt to elucidate the microscopic underpinnings of these and related materials,
recent theoretical studies have produced a general low-energy effective spin-1/2 model for magnetism in rare earth pyrochlores.\cite{Onoda2011,Savary2012,Lee2012}  
In an important development, Huang, Chen and Hermele\cite{Huang2014} 
have shown that, on the pyrochlore lattice, 
strong spin-orbit coupling can lead to Kramers doublets with dipolar-octupolar character in $d$- and $f$-electron systems.
This allows for a specialization of the general effective model to one without the debilitating ``sign problem'' -- amenable to solution 
through quantum Monte Carlo (QMC) methods -- thus admitting a systematic search for QSL phases via large-scale computer simulations.

\section*{\uppercase{A quantum kagome ice model}}
\begin{figure*}[ht]
 \includegraphics[width=0.74\textwidth]{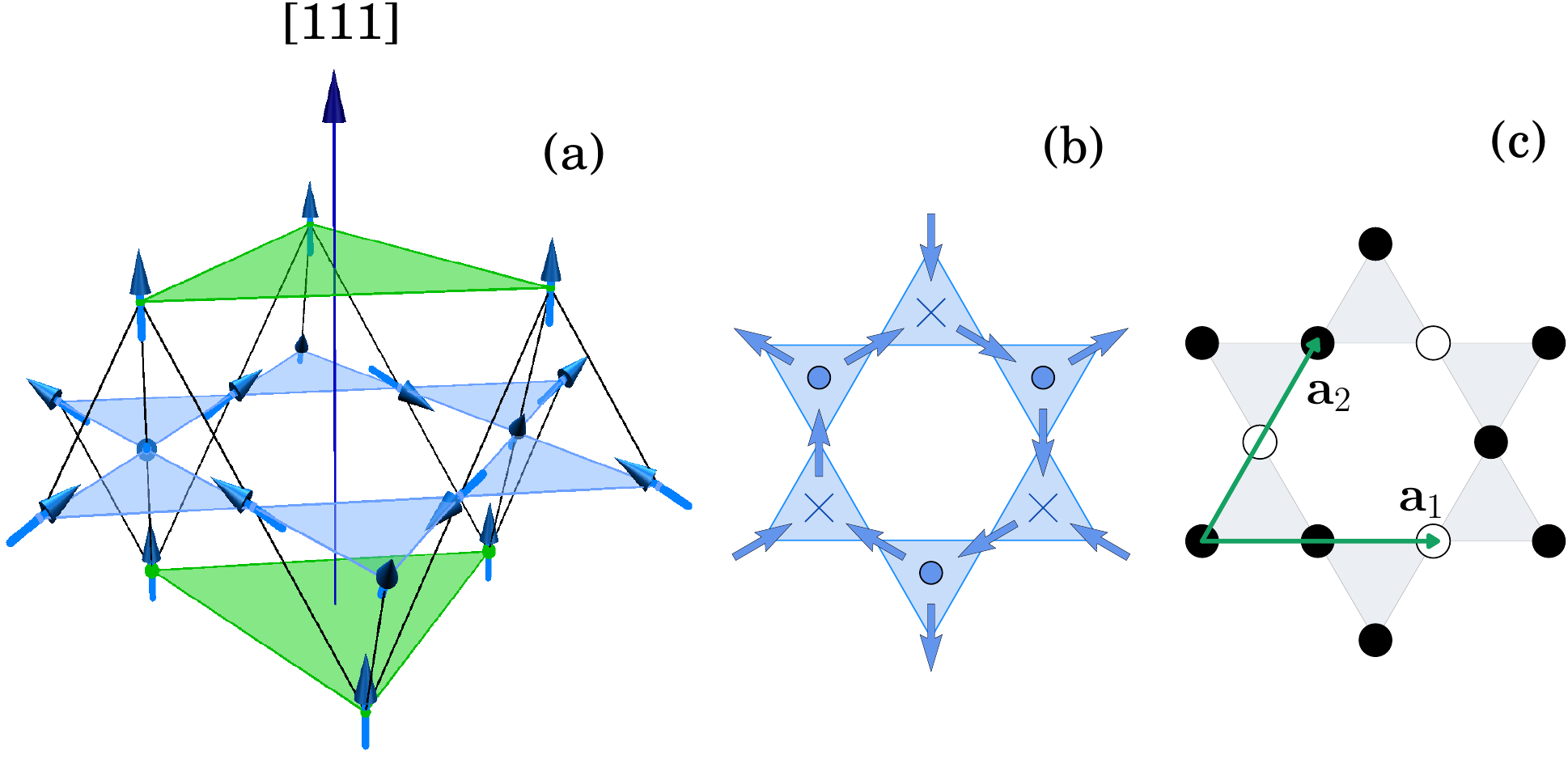}
\caption{\label{fig:mapping} {\bf From pyrochlore spin ice to kagome ice.} (a) The pyrochlore lattice viewed as a set of alternating kagome (blue) and triangular (green) layers along the $[111]$ direction. Spins on the pyrochlore lattice satisfy the ``ice rules'': two-in, two-out of each tetraheron. (b) Two-dimensional projection of the pyrochlore spin configuration onto a kagome plane.
At the center of each triangle is a representation of the out-of-plane spin: dots (crosses) refer to a spin pointing inward (outward) of each tetrahedron in (a).
(c) The associated pseudo-spin $S^z$ configuration of Eq.~(\ref{HAM}) where, filled (empty) circles represents a pseudo-spin up (down). Our QMC simulations of the pseudo-spin Hamiltonian are defined on periodic tori spanned 
by the primitive vectors $\mathbf{a}_1$ and $\mathbf{a}_2$, where $\lVert\mathbf{a}_1\rVert=\lVert\mathbf{a}_2\rVert=2$. } \label{fig:lattice}
\end{figure*}

While the possibility for 3D QSLs in the above compounds is intriguing, 
spin ice materials offer a compelling mechanism for dimensional reduction to 2D,
since single-ion anisotropy constrains magnetic moments to point along the {\it local} tetrahedral symmetry axes in the pyrochlore lattice.
This mechanism consists of the application of an external magnetic field 
along the {\it global} [111] crystallographic direction that partially lifts the spin ice degeneracy by ``pinning'' one spin per tetrahedron. %
As illustrated in Fig.~\ref{fig:lattice}a, this [111] magnetic field effectively decouples spins between the alternating kagome and triangular layers of the original pyrochlore
structure.
To simplest approximation, the system becomes a two-dimensional system of stacked kagome planes,\cite{Matsushira2002,Moessner2003,Isakov2004,Macdonald2011} 
where spins on the intervening triangular planes align in the direction of the field (becoming energetically removed from the problem), 
while those in the kagome plane remain partially disordered.
These kagome spins
retain a fraction of the zero-field spin ice entropy, though 
still preserving the spin ice rules (two-in, two-out) of the parent pyrochlore system. 
This leads to classically disordered state, termed ``kagome ice'',\cite{Matsushira2002,Wills2002,Isakov2004,Macdonald2011} 
evidenced to date in several experimental studies on spin ice materials.\cite{Hiroi2003,Sakakibara2003,Tabata2006}

The above observations lead to a natural microscopic mechanism to search for 2D QSL behaviour.\cite{Molavian2009}
First, one begins with classical nearest-neighbor spin ice in an applied [111] field, so as to promote the aforementioned ``kagome ice'' state.
This model maps to a projected {\it pseudo}-spin Ising model with a symmetry-breaking Zeeman field $h$,
arising from a combination of the physical [111] field and the original pyrochlore spin exchange interaction (Fig.~\ref{fig:lattice}c).
For moderate $h$, the classical ground state retains an extensive degeneracy, before becoming a fully-polarized ferromagnetic state
for $h/J_z > 2$.
Next, to include the effect of quantum fluctuations, one may add exchanges
from the recent quantum spin ice models.\cite{Onoda2011,Lee2012,Savary2012,Huang2014}  
We consider only those quantum fluctuations discussed by Huang {\it et.~al.}\cite{Huang2014}, 
to obtain a pseudo-spin Hamiltonian on the kagome lattice,
\vspace{-1.0 mm}
\begin{eqnarray}
\mathcal{H}_{\text{XYZ}} &=& \sum\limits_{\langle  \boldsymbol{r}  \boldsymbol{r}' \rangle} \Big[ {J}_{z} S^{z}_{\boldsymbol{r}} S^{z}_{\boldsymbol{r}'} 
- { \frac{J_{\pm}}{2}}( S^{+}_{\boldsymbol{r}}   S^{-}_{\boldsymbol{r}'} + h.c.)  \nonumber \\
&+& { \frac{J_{\pm\pm}}{2}}( S^{+}_{\boldsymbol{r}} S^{+}_{\boldsymbol{r}'} + h.c.  ) \Big] -h\sum\limits_{\boldsymbol{r} }S^{z}_{\boldsymbol{r}}  .
\label{HAM}
\end{eqnarray}
Here, ${\mathbf S_r}$ are spin-1/2 operators, with a global $z$ axis ($S^{z}_{\boldsymbol{r}} = 1/2 = \CIRCLE$ and $S^{z}_{\boldsymbol{r}} = -1/2 = \Circle$ in
 Fig.~\ref{fig:lattice}c).
This Hamiltonian cannot be solved exactly by analytical techniques; however large-scale QMC simulations are possible  
in a parameter regime devoid of the prohibitive ``sign problem'', which occurs for $J_{\pm} >0$. 

\begin{figure*}[ht]
\includegraphics[width=0.75\textwidth]{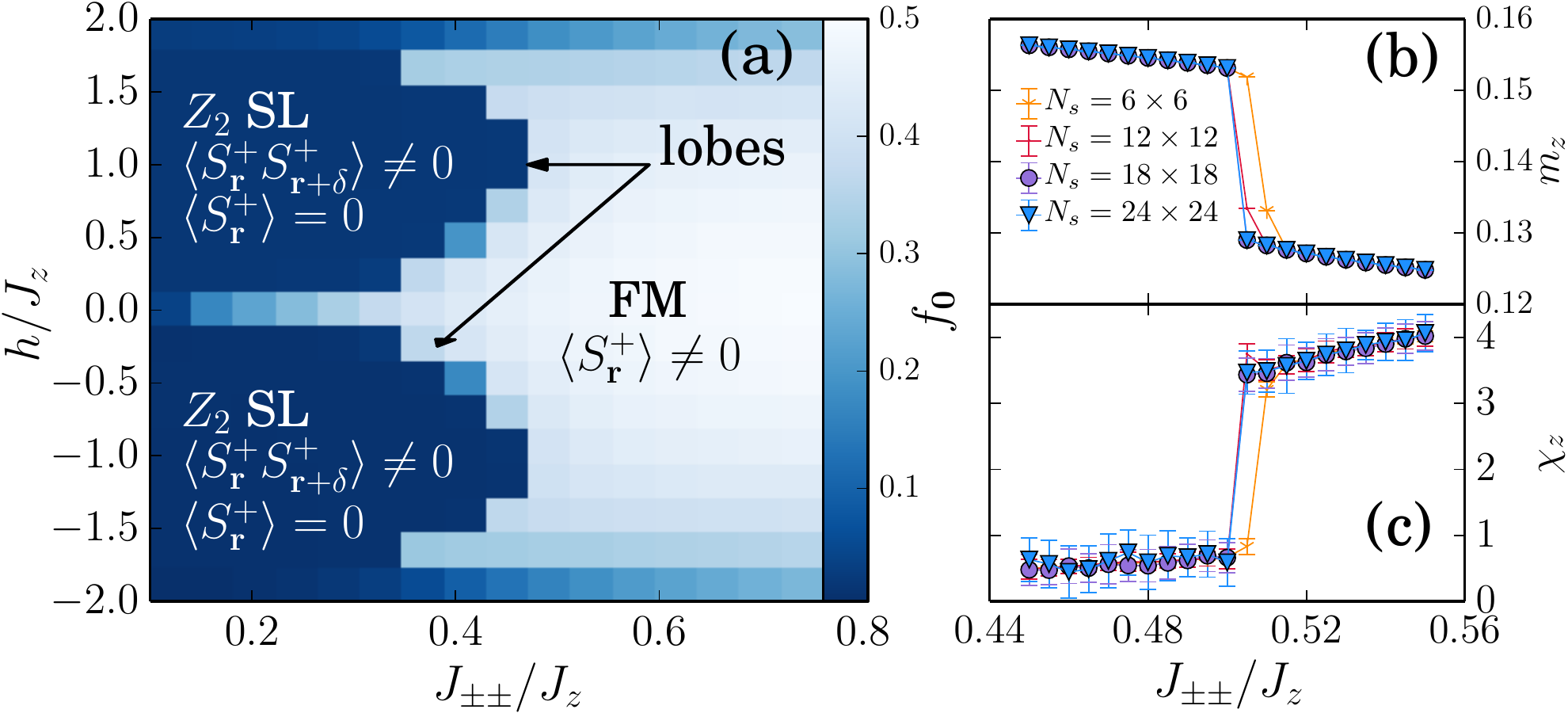}
\caption{\label{fig:phased}{\bf Phase diagram of the model.} (a) Phases of the model in Eq.~\ref{HAM} for $J_{\pm}=0$ as a function of function of $h/J_z$ and
$J_{\pm\pm}/J_z$. The color scale represents the zero momentum occupation $f_0$ obtained from a $N_s=6\times6$ lattice and temperature
$T=J_z/24$. Magnetization (b) and uniform susceptibility (c) as a function of $J_{\pm\pm}/J_z$ entering the $Z_2$ QSL lobe at fixed $h/J_z=0.833$.
Note that a fully spin-polarized phase occurs for $h/J_z \ge 2$, which is not illustrated in this phase diagram.
}
\end{figure*}

One can imagine a 2D QSL state arising conceptually by considering the quantum fluctuations 
$J_{\pm}$ and $J_{\pm \pm}$ as perturbations on the classical kagome ice limit, where only diagonal terms
$J_z>0$ and $h \ll J_z$ are present. Previously, large-scale QMC simulations have been performed on the kagome model in the limit $J_{\pm}>0$ and 
$J_{\pm\pm}=0$,\cite{Isakov2006,Damle2006} (a parameter regime where the Hamiltonian retains U(1) invariance). In that case, quantum fluctuations promote 
an in-plane ferromagnetic (FM) phase for $h=0$, and a ``valence bond-solid'' (VBS -- a conventional symmetry broken phase) for $h>0$.
Thus, it happens that fluctuations of the form induced by $J_{\pm}$ are not sufficient to promote a 2D QSL state.

However, there remains the theoretical
possibility of a gapped $Z_2$ QSL phase promoted by the $J_{\pm \pm}$ quantum fluctuations. 
As detailed in the Supplemental Information,
the local constraints of classical kagome ice can be translated into a charge-free condition on the dual honeycomb lattice.
Then, the full Hamiltonian (\ref{HAM}) can be re-cast as a system of interacting bosonic spinons coupled to a  compact U(1) gauge field on the dual  
lattice. In the limit of $J_{\pm}=0$, this theory is expected to exhibit two distinct phases. One is a ``confined'' phase, corresponding 
to a conventional spin-ordered state; the other is a ``deconfined'' $Z_2$ QSL phase.\cite{Fradkin.1979,Lee2012,Huang2014}  From these simple arguments it is conceivable that these two phases 
exist in the phase diagram of Eq.~(\ref{HAM}).
In the next section, we set $J_{\pm}=0$ and explore this possibility for all parameter regimes 
$J_{\pm \pm}/J_z$ and $h/J_z$, using non-perturbative, unbiased QMC simulations. 

\section*{\uppercase{Quantum monte carlo results}}

We implement a finite-temperature Stochastic Series Expansion~\cite{sandvik1999,Syljuaasen2002,Melko2007} (SSE) QMC algorithm 
with {\it directed} loop updates in a $2+1$ dimensional simulation cell, 
designed specifically to study the Hamiltonian Eq.~(\ref{HAM}) with $J_{\pm}=0$
(for details, see the Methods Summary).
Note, this Hamiltonian explicitly breaks U(1) invariance, retaining global $Z_2$ symmetries. 
By a canonical transformation, $S^{\pm}\to\pm iS^{\pm}$; we simulate only $J_{\pm\pm}<0$, without loss of generality.\cite{Huang2014}
Various measurements are possible in this type of QMC simulation.
Simplest are the standard SSE estimators for energy, magnetization $m_z=\langle \hat{m}\rangle=\langle\frac{1}{V}\sum_iS_i^{z}\rangle$, and uniform spin susceptibility $\chi_z=\frac{V}{T}\left(\langle \hat{m}^{2} \rangle -\langle \hat{m} \rangle^2 \right)$.  The 
latter two allow us to map out the broad features of the phase diagram.   
Further, we measure the off-diagonal spin structure factor~\cite{Dorneich2001} 
\begin{equation} 
\label{momdist}
n^{\alpha\beta}_{\boldsymbol{q}} = \frac{1}{N_s} \sum\limits_{\boldsymbol{r}_i \boldsymbol{r}_j}  e^{i \boldsymbol{q} \left[ \left(  \boldsymbol{r}_i+ \boldsymbol{\alpha}  \right)  -   \left(  \boldsymbol{r}_j+ \boldsymbol{\beta}  \right)     \right]} \langle S^{+}_{ \boldsymbol{r}_i+ \boldsymbol{\alpha} } S^{-}_{ \boldsymbol{r}_j+ \boldsymbol{\beta} }\rangle.
\end{equation}
Here, $\boldsymbol{r}_i$ points to the sites of the underlying triangular lattice (containing $N_s$ sites) of the kagome lattice (containing $V=3\times N_s$ sites).
The vectors $\boldsymbol{\alpha}$ are the position of each site within the unit cell with respect to the vector $\boldsymbol{r}_i$. This quantity allows us to define, for this spin Hamiltonian, the 
analogue of a  condensate fraction in bosonic systems,\cite{Penrose1956,Giamarchi2008} which detects transverse magnetic ordering. We define $f_0=\frac{n_{M}}{V}$ as the 
ratio of largest eigenvalue $n_M$ of the ``single-particle'' density matrix $\rho_{i,j}=\langle S^{+}_{\boldsymbol{r}_i } S^{-}_{\boldsymbol{r}_j} \rangle$ to the number 
of sites $V$. The eigenvalues of $\rho_{i,j}$ coincide with $n_{\boldsymbol{q}}=\sum\limits_{\alpha} n^{\alpha,\alpha}_{\boldsymbol{q}}$ for a translationally invariant system. 

Figure \ref{fig:phased} shows the QMC phase diagram for the $J_{\pm}=0$ model of Eq.~(\ref{HAM}), 
using data for the condensate fraction $f_0$. Careful finite-temperature and finite-size scaling, performed up to lattice sizes of 
$V = L \times L \times 3 = 39 \times 39 \times 3$ and $\beta = J_{z}/T = 96$, is detailed in the Supplemental Information. The magnetization curve and
the uniform spin susceptibility across the phase boundary at fixed $h/J_z=0.833$ are presented in Fig.~\ref{fig:phased}.
The data clearly indicates the existence of two magnetized ``lobes'' on the phase diagram for $J_{\pm\pm}/J_z < 0.5$ and $h/J_z \neq 0$,
where the zero-momentum condensate fraction of a surrounding FM phase is destroyed
by a phase transition (which appears to be first order).
The lobes have magnetizations of $ m  \approx  - 1/6$ and $ m  \approx  +1/6$  for $h/J_z < 0$ and $h/J_z > 0$, respectively. 
The FM phase has a finite uniform susceptibility $\chi_z$,
while the lobe phases retain a small but finite $\chi_z$ that can be understood by the nature of the
quantum fluctuation $( S^{+}_{\boldsymbol{r}}S^{+}_{\boldsymbol{r}'} + S^{-}_{\boldsymbol{r}} S^{-}_{\boldsymbol{r}'})$ as a spin {\it pair} interaction, which does not conserve the total magnetization $S^{z}_{tot}$.
As discussed above, the phase in these lobes is a candidate for supporting a 2D QSL state.

\begin{figure*}[t]
  \includegraphics[trim=0 0 0 0, clip,width=0.8\textwidth]{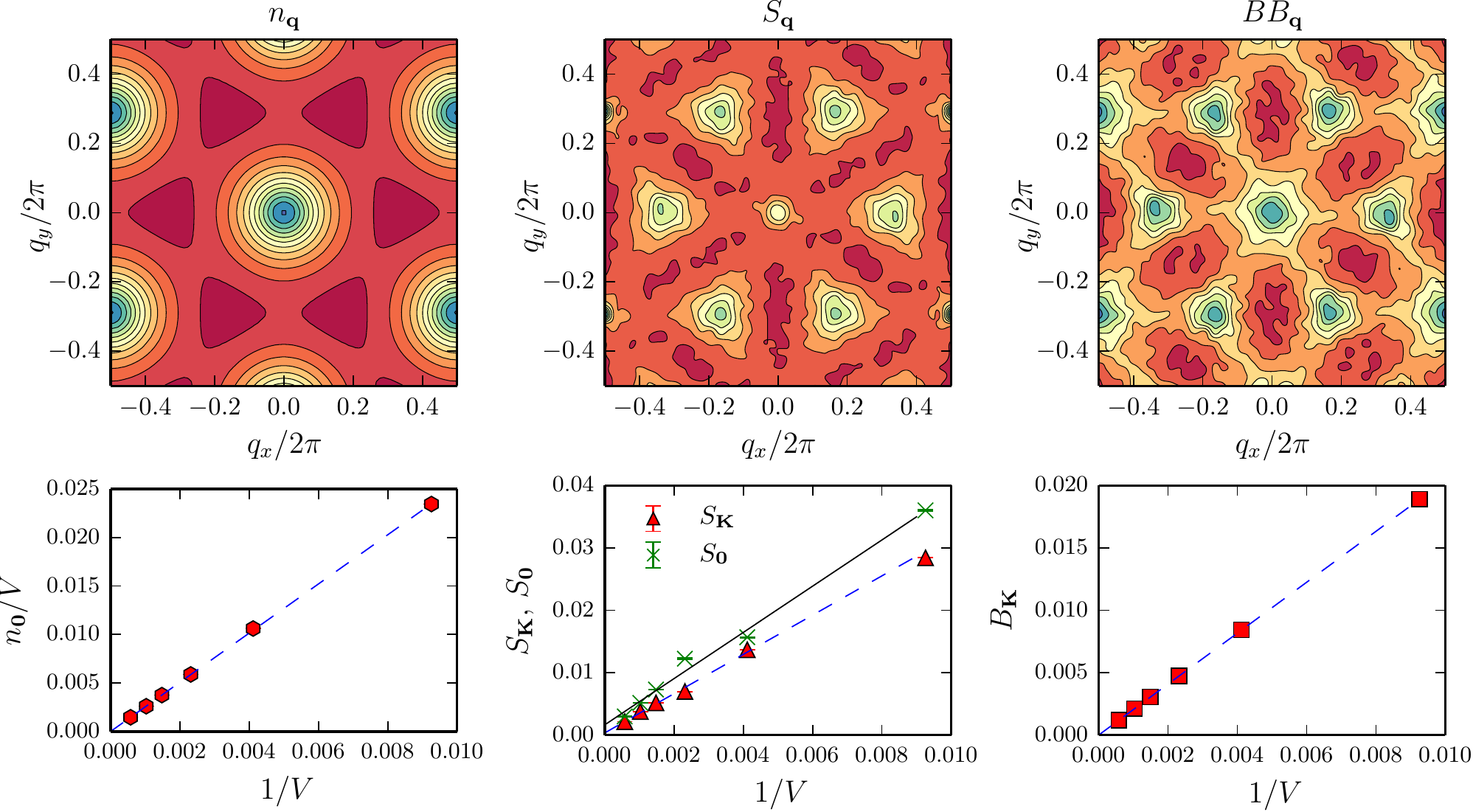}
  \caption{{\bf Structure factors and absence of order in the lobe.} Top: Off-diagonal $n_{\boldsymbol{q}}$, diagonal  $S_{\boldsymbol{q}}$, and bond $BB_{\boldsymbol{q}}$ structure factors inside the lobe for a system with $N_s=24\times24$, $h/J_z=0.8333$, $J_{\pm\pm}/J_z=0.495$, and $T=J_z/48$. Bottom: The corresponding finite-size scaling of candidate peaks at ${\bf q}$ values where local maxima occur in the structure factor.\label{fig:sfactors} The zero-momentum peak of $BB_{\boldsymbol{q}}$ has been removed.}
\end{figure*}
In order to examine this hypothesis, we perform a detailed search for ordered structures in the lobes.
In related models, particularly the spin-1/2 XXZ model on kagome (i.e.~$J_{\pm\pm}=0$ and $J_{\pm} >0$),\cite{Isakov2006,Damle2006} the analogous lobes support 
a conventional VBS phase, which is evident in the diagonal structure factor:
$S^{\alpha\beta}_{\boldsymbol{q}}/N_s =\langle S^{\alpha}_{\boldsymbol{q}}  S^{\beta}_{\boldsymbol{-q}} \rangle -\langle S^{\alpha}_{\boldsymbol{q}}\rangle \langle S^{\beta}_{\boldsymbol{-q}} \rangle$, 
where  
\begin{equation}
S^{\alpha}_{\boldsymbol{q}}=\frac{1}{N_s} \sum\limits_{\boldsymbol{r}_i} e^{ i\boldsymbol{q}\left(  \boldsymbol{r}_i+ \boldsymbol{\alpha}\ \right) } S^{z}_{\boldsymbol{r}_i+ \boldsymbol{\alpha}}.
\end{equation}
If there there is long-range order then $S_{\boldsymbol{q}}=\sum_{\alpha}S^{\alpha\alpha}_{\boldsymbol{q}}$ will scale with system size for at least one value of $\boldsymbol{q}$. We also measure the bond-bond structure factor using a four-point correlation function.
\begin{equation}
BB^{\alpha\beta}_{\boldsymbol{q}}=\frac{1}{N_s}\sum_{\boldsymbol{r}_a\boldsymbol{r}_b} e^{i\boldsymbol{q}\left( \boldsymbol{r}_a -\boldsymbol{r}_b \right) }\langle B^{\alpha}_{\boldsymbol{r}_a} B^{\beta}_{\boldsymbol{r}_b} \rangle,
\end{equation}
 where $B^{\alpha}_{\boldsymbol{r}_a}=S_{i_{a\alpha}}^{+}S_{j_{a\alpha}}^{+}+S_{i_{a\alpha}}^{-}S_{j_{a\alpha}}^{-}$. 
Nearest neighbor sites $i_{a\alpha}$ and $j_{a\alpha}$ belong to bond $\alpha$ in a unit cell located at position $\boldsymbol{r}_a$. Again, if there is pair long-range order then $BB_{\boldsymbol{q}}=\sum_{\alpha}BB^{\alpha \alpha}_{\boldsymbol{q}}$ should scale with system size for at least one value of $\boldsymbol{q}$, with which we define $B_{\boldsymbol{q}}=BB_{\boldsymbol{q}}/V$. 

Figure \ref{fig:sfactors} illustrates the various ${\bf q}$-dependent structure factors for spin and bond order.  
These structure factors display diffuse peaks at various wave vectors, notably $\boldsymbol{q}=\boldsymbol{0}$, $\boldsymbol{q}=\boldsymbol{K}=(2\pi/3,0)$, 
and symmetry-related momenta.  Such peaks would indicate the presence of long-range order, should they sharpen, and survive in intensity in the infinite-size limit, where $S/V$ would 
correspond to an order parameter squared. In the insets to Fig.~\ref{fig:sfactors}, we examine this through a standard finite-size scaling analysis, for several
candidate peaks for each of the structure factors. Further scaling analysis, including larger system sizes, is presented in the Supplemental Information.
In each case, the QMC data indicates a scaling of each order parameter to zero in the limit $V \rightarrow \infty$. 
Note in particular, the largest value of  $B_{\boldsymbol{q}}$ corresponds to $\boldsymbol{q}=0$, which remains finite as $V\to\infty$, meaning that the bond expectation values $\langle S_{i_{a\alpha}}^{+}S_{j_{a\alpha}}^{+} \rangle\neq0$ is finite in the lobes. This is expected as $\langle S_{i_{a\alpha}}^{+}S_{j_{a\alpha}}^{+} \rangle$ represents the ``kinetic energy'' of quantum fluctuations in the system, thus it should be finite in all phases. 
More importantly, the data indicates that in the limit of $V\to\infty$ this quantity is the same on all bonds of the unit cell of the kagome lattice, meaning that there is no breaking of space-group symmetry (see Supplemental Information).

\begin{figure}[b]
  \includegraphics[width=0.4\textwidth]{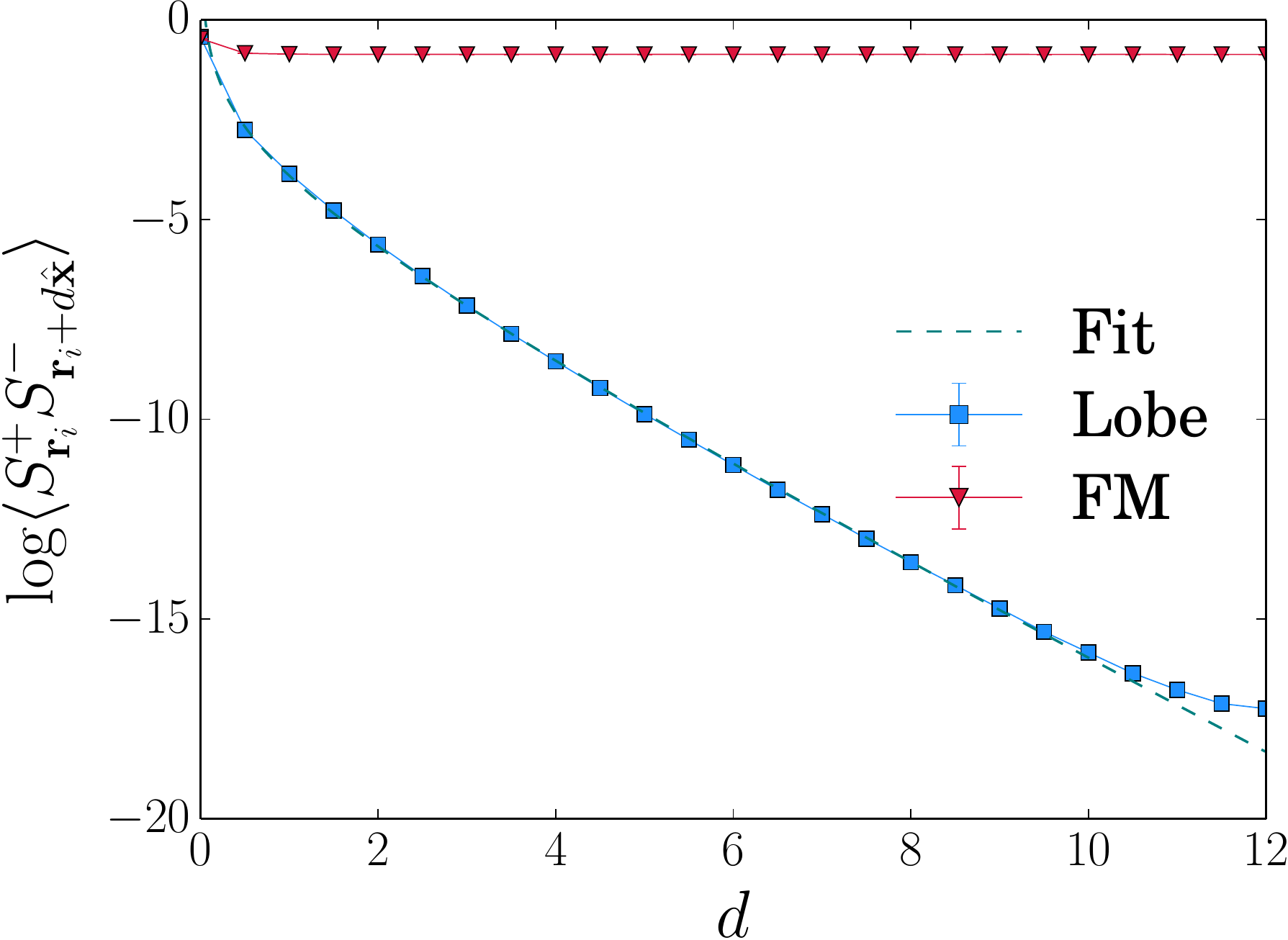}
  \caption{\label{fig:corrf} {\bf Exponential decay of correlation functions.} The off-diagonal spin correlation function as a function of distance along the $\hat{\mathbf{x}}$ direction for a system with $N_s=24\times24$ and $T=J_z/48$. The dashed line corresponds to a fit of the numerical data  to the function 
$f\left(d\right)=c-\frac{d}{\xi}-\alpha\ln d$}
\end{figure}

Finally, since the above data suggests the existence of a phase that is homogeneous, disordered, and quantum-mechanically fluctuating at
extremely low temperatures, one should also examine whether the energy for excitations out of this ground state is gapped or gapless.  
Although a direct measurement of the gap is not possible in this type of SSE QMC method, we can indirectly probe its existence by
looking at the decay of real-space correlations. In Fig.~\ref{fig:corrf}, we compare the decay of single-particle correlations between the 
$m_z = \pm 1/6$ magnetization lobes, and the adjacent FM ordered phase.
For the system size studied, it is clear that that correlations in the lobe are consistent with exponential decay,
and therefore indicative of a gap. In contrast, in the FM phase the correlations quickly reach a  finite value, indicating symmetry breaking. 
Similarly, the diagonal part of the spin correlation function is consistent with exponential decay both in the lobes and in the FM phase (not illustrated).

Thus, our QMC results have elucidated a phase diagram for our kagome pseudo-spin XYZ model (with $J_{\pm}=0$)
which contains a predominant FM phase, surrounding ``lobes'' of  an exotic disordered $m_z = \pm 1/6$ magnetization phase.  
Since our dual gauge theory (detailed in the Supplemental Information) indicates that these lobes may realize a QSL with an 
emergent $Z_2$ gauge symmetry, it is clear that further simulation work should be carried out to address this hypothesis.
To confirm the presence of a $Z_2$ QSL, one requires evidence of either excitations consistent with this gauge structure
(e.g.~magnetic spinons or non-magnetic visons at non-zero temperature), or a smoking gun such as the 
topological entanglement entropy.\cite{KP,LW}  Such evidence, although demonstrated in the past with SSE QMC,\cite{Isakov2011,Tang2011}
is resource-intensive to obtain, requiring high numerical precision at very low temperatures,
and thus outside of the scope of the present manuscript.

However, we also note that, due to the presence of only a discrete symmetry in our kagome $XYZ$ model, an emergent $Z_2$
structure is not strictly
required by the Lieb-Schultz-Mattis-Hastings (LSMH) theorem,\cite{Oshikawa2000,Hastings2004,Nachtergaele2007}
which states that a system with half-odd-integer spin in the unit cell cannot have a gap {\it and} a unique ground state.
In higher-symmetry Hamiltonians, the requirements of the LSMH theorem are satisfied in a gapped QSL
phase by the {\it topological} degeneracy, which is a consequence of the emergent discrete gauge symmetry.
For our Hamiltonian with a gapped QSL arising in a model with only global discrete symmetries, 
an emergent gauge structure is not required.  Rather, it is possible 
that the groundstate is a quantum paramagnet. 
On the other hand,  
other types of emergent gauge structure, topological order, or
other exotic phenomena are theoretically possible.
Fortunately, the nature of this Hamiltonian, which is among the first to show a 2D QSL phase with only nearest-neighbor interactions, lends itself 
exceedingly well to study by sign-problem-free QMC simulations. We therefore expect a large number of studies in the near future will help elucidate the precise nature of this QSL phase.  

\section*{\uppercase{Conclusion and outlook}}

Through extensive quantum Monte Carlo (QMC) simulations, we have studied a sign-problem-free model of frustrated quantum spins 
interacting on a two-dimensional (2D) kagome lattice.  This model is decedent from a more general quantum XYZ Hamiltonian discussed by Huang, Chen and Hermele,\cite{Huang2014}
derived for the three-dimensional pyrochlore lattice, when subject to a magnetic field along the [111] crystalographic direction.
For a large range of Hamiltonian parameters, the QMC data uncovers an exotic disordered phase 
which breaks no symmetries, has strong quantum mechanical fluctuations and exponentially decaying correlations
-- a gapped quantum spin liquid (QSL) phase.
This discovery is consistent with an analytical dual gauge theory (detailed in the Supplemental Information), 
which indicates that, in the limit of small quantum fluctuations, the phase could be a 2D QSL with an emergent $Z_2$ gauge symmetry. 

Our work suggests a new experimental avenue to search for the highly-coveted QSL phase in two dimensions.  Previous efforts have focussed largely on 
SU(2) Hamiltonains on kagome or triangular lattice materials.\cite{Pratt2011,Han2012}   In contrast, we propose to concentrate
the search on the quantum spin ice pyrochlore materials, subject to an external field along the [111] direction.  
Such kagome ice phases have been identified in various materials in the past.  A closer look at several quantum spin ice candidates is warranted, 
particularly in materials where strong quantum fluctuations are known to exist, such as Tb$_2$Ti$_2$O$_7$,  Yb$_2$Ti$_2$O$_7$, Pr$_2$Zr$_2$O$_7$, and Pr$_2$Sn$_2$O$_7$.
Also, in light of recent experiments\cite{Pomaranski2013} which suggest that the oft-studied classical spin ice state is only metastable in Dy$_2$Ti$_2$O$_7$,
it would seem prudent to re-examine the kagome ice state of this material using similar long-timescale techniques, to ascertain whether  
evidence of a QSL state may be present yet dynamically inhibited in the short-timescale studies performed to date.

\section*{Methods Summary}
We developed a Stochastic Series Expansion~\cite{sandvik1999} (SSE) QMC algorithm in the global $S^z$ basis
designed to study the Hamiltonian Eq.~(\ref{HAM}) with $J_{\pm}=0$ at finite temperature, using a $2+1$-dimensional simulation cell.
Within the SSE, the Hamiltonian was implemented with a triangular plaquette breakup,\cite{Melko2007} which helps ergodicity in the regime where $J_z/J_{\pm \pm}$ is large.
Using this Hamiltonian breakup, the standard SSE {\it directed loop} equations\cite{Syljuaasen2002}
were modified to include sampling of off-diagonal operators of the type $S^{+}_{\boldsymbol{r}} S^{+}_{\boldsymbol{r}'} + h.c$.
The resulting algorithm is highly efficient, scaling linearly in the number of lattice sites $V$ and the inverse temperature $\beta$.
This scaling is modified to $V^2 \beta$ in the cases where a full ${\bf q}$-dependent structure factor measurement is required.

The program was implemented in Fortran and verified by comparing results for small clusters with exact
diagonalization data. For each set of parameters in Eq.~(\ref{HAM}), the simulation typically requires 
$10^{7}$ QMC steps, with approximately 10\% additional equilibration steps. The data presented in this paper required computational resources equivalent to 
100 CPU core-years, run on a high-performance computing (HPC) cluster with Intel Xeon CPUs running at 2.83 GHz clock speed. 


\acknowledgements
We would like to thank F. Becca, A. Burkov, L. Cincio, T. Senthil, and M. Stoudenmire for enlightening discussion, and M. Gingras for a critical reading of the manuscript. We are
particularly indebted to Gang Chen for bringing the models discussed in Ref.~\onlinecite{Huang2014} to our attention and for stimulating 
discussions motivating this study. This research was supported by NSERC of Canada, the Perimeter Institute for Theoretical Physics, and the John Templeton Foundation. 
R.G.M.~acknowledges support from a Canada Research Chair.
Research at Perimeter Institute is supported through Industry Canada
and by the Province of Ontario through the Ministry of Research \& Innovation. 
Numerical simulations were carried out on the Shared Hierarchical Academic Research Computing Network
(SHARCNET).

\section*{\uppercase{Supplemental Information}}
\subsection*{Effective gauge theory}
To understand the possible phases of the Hamiltonian \eqref{HAM}, we concentrate on its classical limit first:
\begin{equation}\label{supp:hclassical}
\mathcal{H}=J_z\sum_{\langle \boldsymbol{r}\boldsymbol{r}^\prime\rangle}S^{z}_{\boldsymbol{r}}S^{z}_{\boldsymbol{r}^\prime}-h\sum_{\boldsymbol{r}}S_{\boldsymbol{r}}^z. 
\end{equation}
At $h=0$, the energy is minimized by any spin configurations with two spins up (down) one spin down (up) on every triangle. The number of such spin configurations increases exponentially with the system size. A finite magnetic field $h>0$ \emph{partially} lifts the extensive degeneracy: the energy of \eqref{supp:hclassical} is minimized by any spin configuration with two spin pointing up and one spin pointing down on each triangle\cite{Moessner2000}. The ground state degeneracy remains extensive. It has been well established that that the manifold of such states can be described by electric flux configurations with no charges present \cite{Damle.2006}. 

\begin{figure}[t]
 \includegraphics[trim=0 0 0 0, clip,width=0.25\textwidth]{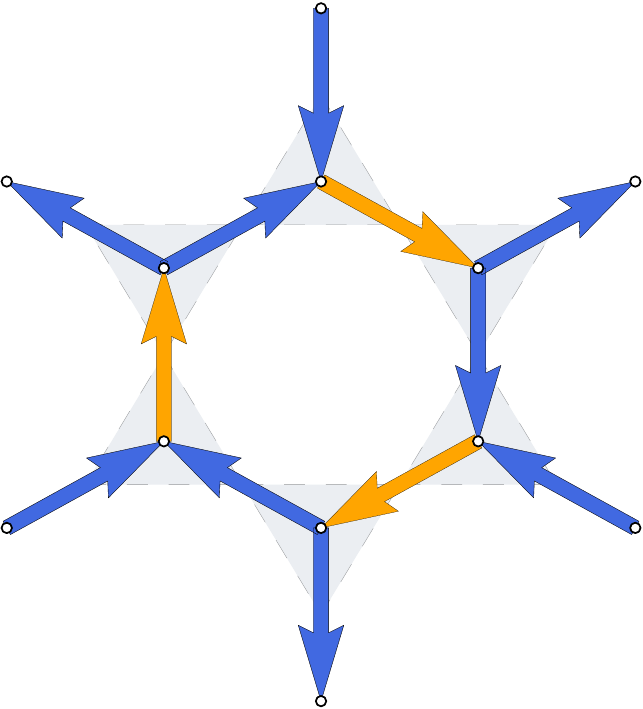}
\caption{\label{fig:divfree} Divergence-free gauge field configuration on the honeycomb lattice corresponding to the pseudo-spins configuration in Fig.\ref{fig:mapping}(c). 
Blue (orange) arrows indicate electric flux of $1/3$  ($2/3$). } 
\end{figure}

To make the assertion explicit, we use $\boldsymbol{x}$ to label the centers of triangles forming a honeycomb lattice. $\boldsymbol{r}$ labels sites on the kagome lattice. A nearest neighbor bond $\langle \boldsymbol{x}\boldsymbol{x}^\prime\rangle$ connecting sites $\boldsymbol{x}$ and $\boldsymbol{x}^\prime$ on the honeycomb lattice goes through site $\boldsymbol{r}$ on the kagome lattice.  We translate $S^{z}_{\boldsymbol{r}}$ to electric flux $E_{\boldsymbol{x},\boldsymbol{x}^\prime}$ using the following relation (see Fig.\ref{fig:mapping}(c) and Fig.\ref{fig:divfree}):
\begin{equation}\label{supp:translation}
E_{\boldsymbol{x},\boldsymbol{x}+\eta_{\boldsymbol{x}}\hat{\mu}}=\eta_{\boldsymbol{x}}\left[\frac{1}{6}-S_{\boldsymbol{r}}^z\right]
\end{equation}
where $\eta_{\boldsymbol{x}}=\pm 1$ if site $\boldsymbol{x}$ belongs to A or B sublattice of the honeycomb lattice. $\hat{\mu}$ ($\mu=1,2,3$) are the three vectors connecting site $\boldsymbol{x}$ with its nearest neighbors. $\hat{1}=(0,1)$, $\hat{2}=(\sqrt{3}/2,-1/2)$ and $\hat{3}=(-\sqrt{3}/2,-1/2)$. The total charge on site $\boldsymbol{x}$ satisfies the Gauss's law:
\begin{equation}\label{supp:gauss}
Q_{\boldsymbol{x}}=\sum_{\mu=1}^{3}E_{\boldsymbol{x},\boldsymbol{x}+\eta_{\boldsymbol{x}}\hat{\mu}}. 
\end{equation}
Use these relations, we rewrite \eqref{supp:hclassical} as:
\begin{equation}\label{supp:translation}
\mathcal{H}=\sum_{\boldsymbol{x}}\left[\frac{J_z}{2}Q_{\boldsymbol{x}}^2-\frac{J_z-h}{2}\eta_{\boldsymbol{x}}Q_{\boldsymbol{x}}\right].
\end{equation}
By construction $Q_{\boldsymbol{x}}$ is constrained to take only integer values. For $h>0$, it is now explicit that the ground states of the classical model \eqref{supp:hclassical} correspond to divergence-free electric field configurations, or $Q_{\boldsymbol{x}}=0$.

We now turn our attention to transverse Hamiltonian terms. A transverse operator, $S_{\boldsymbol{r}}^+$ for example, creates a pair of positive and negative charges. We follow Savary and Balents\cite{Savary2012} to introduce conjugate operator $\phi_{\boldsymbol{x}}$: $[\phi_{\boldsymbol{x}},Q_{\boldsymbol{x}^\prime}]=i\delta_{\boldsymbol{x}\boldsymbol{x}^\prime}$. $S_{\boldsymbol{r}}^+$ is mapped to:
\begin{equation}\label{supp:map}
S_{\boldsymbol{r}}^+=\mathrm{e}^{i\eta_{\boldsymbol{x}}(\phi_{\boldsymbol{x}}-\phi_{\boldsymbol{x}+\hat{\mu}})}s_{\boldsymbol{x}\mu}^+\equiv \psi_{\boldsymbol{x}}^{\eta_{\boldsymbol{x}}}s_{\boldsymbol{x}\mu}^+\psi_{\boldsymbol{x}+\eta_{\boldsymbol{x}}\hat{\mu}}^{-\eta_{\boldsymbol{x}}}.
\end{equation}
Here $\psi_{\boldsymbol{x}}^{\pm}$ increases or decreases $Q_{\boldsymbol{x}}$ by $1$ and $s^{+}_{\boldsymbol{x}\mu}=s^{+}_{\boldsymbol{x}+\hat{\mu},\mu}$. Using \eqref{supp:translation}, and \eqref{supp:map}, we write \eqref{HAM} as:
\begin{eqnarray}\label{supp:ham}
&&\nonumber\mathcal{H}=\sum_{\boldsymbol{x}}\left[ \frac{J_z}{2}Q_{\boldsymbol{x}}^2-\frac{J_z-h}{2}\eta_{\boldsymbol{x}}Q_{\boldsymbol{x}}\right.-\\
&&\left.\frac{J_{\pm}}{2}\sum_{\mu<\nu}\left(\psi_{\boldsymbol{x}}^{\eta_{\boldsymbol{x}}}s_{\boldsymbol{x}\mu}^+s_{\boldsymbol{x}+\eta_{\boldsymbol{x}}\hat{\mu}\nu}^-\psi_{\boldsymbol{x}+\eta_{\boldsymbol{x}}(\hat{\mu}-\hat{\nu})}^{-\eta_{\boldsymbol{x}}}+h.c\right)-\right.\\
&&\nonumber\left.\frac{J_{\pm\pm}}{2}\left(\psi_{\boldsymbol{x}}^{\eta_{\boldsymbol{x}}}\psi_{\boldsymbol{x}}^{\eta_{\boldsymbol{x}}}s^+_{\boldsymbol{x}\mu}s^+_{\boldsymbol{x}\nu}\psi_{\boldsymbol{x}+\eta_{\boldsymbol{x}}\hat{\mu}}^{-\eta_{\boldsymbol{x}}}\psi_{\boldsymbol{x}+\eta_{\boldsymbol{x}}\hat{\nu}}^{-\eta_{\boldsymbol{x}}}+h.c\right)
\right].
\end{eqnarray}
The Hamiltonian is invariant under a local $U(1)$ gauge transformation:
\begin{subequations}\label{sup:gaugeT}
\begin{eqnarray}
\psi_{\boldsymbol{x}}^{\pm}&\to& \psi_{\boldsymbol{x}}^{\pm}\mathrm{e}^{\mp i\alpha_{\boldsymbol{x}}},\\
s^{+}_{\boldsymbol{x}\mu} &\to&s^{+}_{\boldsymbol{x}\mu} \mathrm{e}^{i\eta_{\boldsymbol{x}}(-\alpha_{\boldsymbol{x}}+\alpha_{\boldsymbol{x}+\eta_{\boldsymbol{x}}\hat{\mu}})}.
\end{eqnarray}
\end{subequations}
The azimuthal angle of pseudo spin $s_{\boldsymbol{r}}$ is the vector gauge field $0\le A_{\boldsymbol{x}\mu}<2\pi$: the gauge theory is compact. 

Equation \eqref{supp:ham} is our main result. The low energy physics of the spin model \eqref{HAM} is thus described by bosonic matter interacting with a $U(1)$ compact gauge field. The theory possesses two types of ground states.\cite{Fradkin.1979} The first class consists of confined phases where either matter fields are confined or charge $1$ particles are condensed. In terms of spins, such phases correspond to long-range ordered phases where some symmetry of the Hamiltonian is broken. For the other category of ground states, charge $n$ ($n>1$) excitations are condensed. The resulting phase, the so-called charge-$n$ Higgs phase, does not need to break any symmetry of the Hamiltonian, possesses local $Z_n$ gauge structure and intrinsic topological order.  These are the $Z_n$ spin liquids. The simplest example is the $Z_2$ spin liquid with $n=2$. In such a state, the spinon pairing introduces a mass to the gauge field, much like the condensation of cooper pairs in a superconductor generates a mass to the gauge field. This means that, at low energies, the local gauge redundancy reduces from $U(1)$ to $Z_2$ through the Anderson-Higgs mechanism. Both the gauge field and the spinons should be gapped, which is corroborated in our simulations in the sense that both diagonal (gauge field) and off-diagonal (spinon) spin correlation functions are consistent with an exponential decay as a function of distance.  

A simple mean-field decoupling of the spinon interaction term in Eq.~\eqref{supp:ham} for $J_{\pm}=0$ provides further insight into the properties of possible phases. We decompose the four-rotor term in both the hopping channel and the pairing channel\cite{Lee2012,Huang2014} through the following mean-field parameters:
\begin{subequations}\label{supp:ansatz}
\begin{eqnarray}
g&\equiv&\langle \psi_{\boldsymbol{x}}^+\psi_{\boldsymbol{x}+\eta_{\boldsymbol{x}}\hat{\mu}}^{-}\rangle,\\
n&\equiv&\langle\psi_{\boldsymbol{x}}^+\psi_{\boldsymbol{x}}^+\rangle=\langle\psi_{\boldsymbol{x}}^-\psi_{\boldsymbol{x}}^-\rangle,\\
f&\equiv&\langle\psi_{\boldsymbol{x}+\eta_{\boldsymbol{x}}\hat{\mu}}^+\psi_{\boldsymbol{x}+\eta_{\boldsymbol{x}}\hat{\nu}}^+\rangle=\langle\psi_{\boldsymbol{x}+\eta_{\boldsymbol{x}}\hat{\mu}}^-\psi_{\boldsymbol{x}+\eta_{\boldsymbol{x}}\hat{\nu}}^-\rangle.
\end{eqnarray}
\end{subequations}
A finite $g$ indicates semi-classical long-range order since $\langle S^+\rangle = \langle s^+\rangle \langle \psi^\dagger_{\boldsymbol{x}} \psi_{\boldsymbol{x}+\eta_{\boldsymbol{x}}\hat{\mu}}\rangle\sim g$; this is the FM phase obtained in our numerical simulations. On the 
other hand, finite $n$ and $f$ with $g=0$ represents a state with condensed double charged matter fields. It is a gapped $Z_2$ spin liquid state.\cite{Fradkin.1979,Lee2012,Huang2014}
We find that in the lobes $\langle S_{\boldsymbol{r}}^+\rangle\sim g$ vanishes while $\langle S_{\boldsymbol{r}}^{+}S_{\boldsymbol{r'}}^{+}\rangle \sim nf$ remains finite, in agreement with a $Z_2$ spin liquid phase.

\subsection*{Absence of breaking of rotational invariance}
\begin{figure}[t]
  \includegraphics[width=0.48\textwidth]{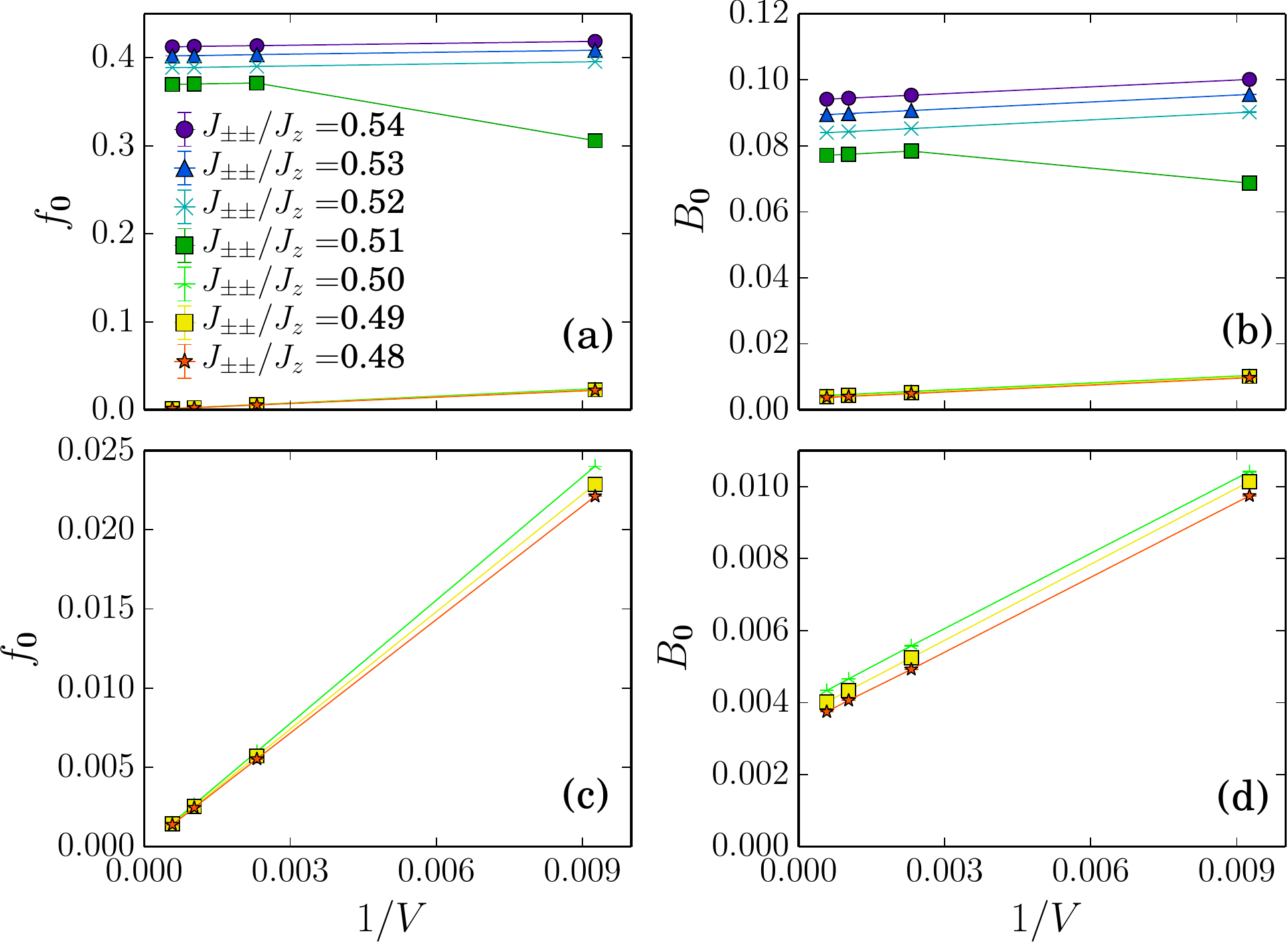}
  \caption{\label{fig:scalacross}(a) Finite-size scaling of the ``single-particle'' zero-momentum occupation $f_0$ across the phase transition between the
ferromagnet and the disordered phases in the lobes. (b) Same as in (a) but for the bond zero-momentum bond occupation $B_0$. Figures (c) and (d) zoom in the size scalings
 in the lobes for $f_0$ and $B_0$, respectively. The temperature is set to $T=J_z/24$.}
\end{figure}

\begin{figure}[t]
  \includegraphics[width=0.42\textwidth]{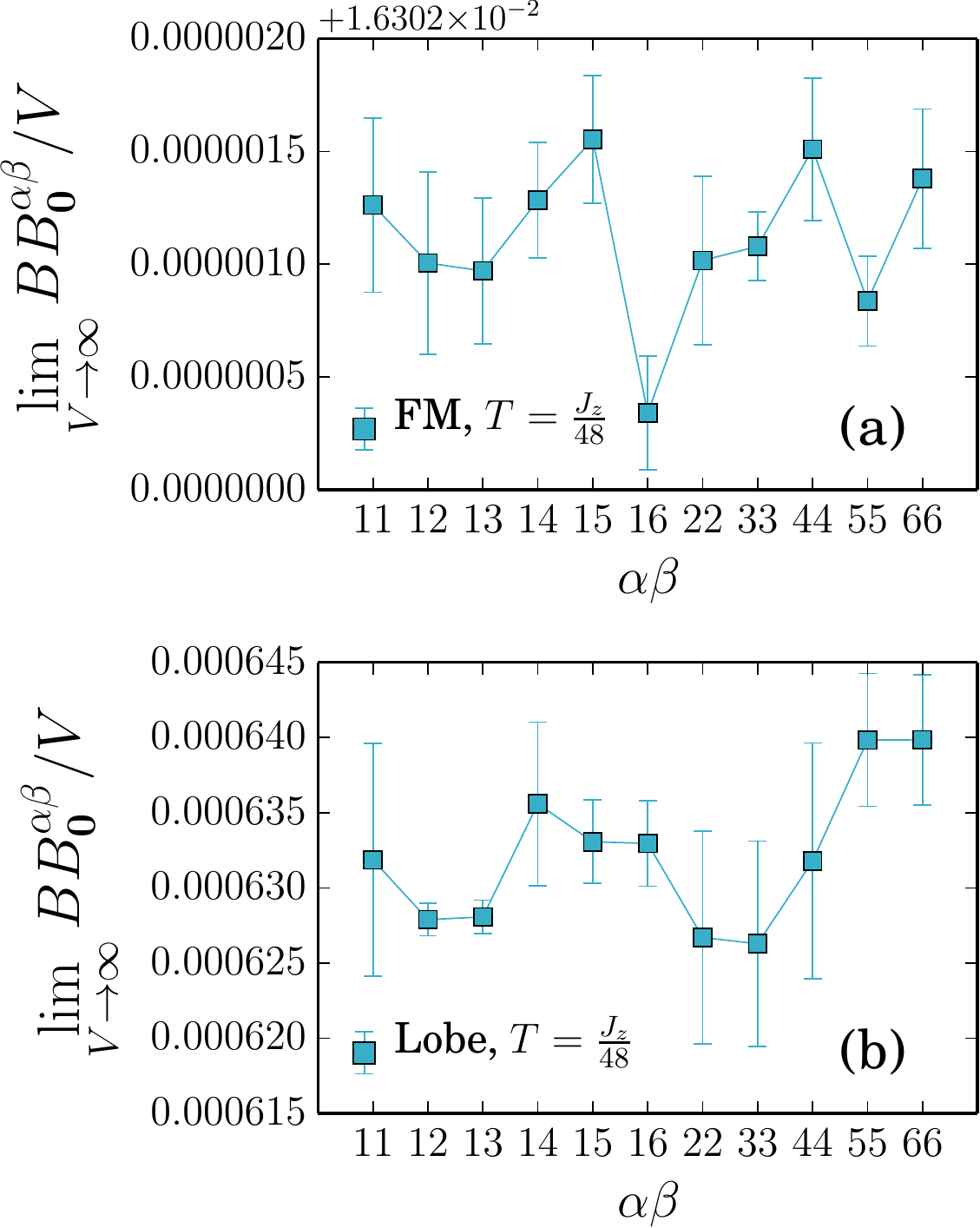}
  \caption{\label{fig:EXTrotinv}The limit of $V\to\infty$ of several matrix elements $BB^{\alpha\beta}_{0}$ as a function of the matrix indices $\alpha\beta$ in the FM phase and
in the lobes ((a) and (b), respectively).}
\end{figure}

In Fig.~\ref{fig:scalacross} we show a finite-size scaling of both the order parameter $f_0$, as well as the zero-momentum bond occupation $B_0$, across
the transition between the FM phase and the disordered phase in the lobes for $T=J_z/24$. Outside the lobes both $f_0$ and $B_0$ remain finite as $V\to\infty$, whereas in the
lobes $f_0$ vanishes (see Fig.~\ref{fig:scalacross}(c)) but $B_0$ remains finite (see Fig.~\ref{fig:scalacross}(d)). Notice that because $B_0$ is finite, 
the state in the lobes can still break the rotational symmetry of the lattice by having different expectation values of $\langle B^{\alpha}_{\boldsymbol{r}_a} \rangle$ over
the 6 independent bonds per unit cell in the kagome lattice. This potential symmetry breaking can be quantified by analyzing the size scaling of the different 
sublattice occupations $B^{\alpha\beta}_{0}=BB^{\alpha\beta}_{0}/V$. In the thermodynamic limit, they are all expected to be equal if the system does not break 
rotational symmetry.  
\begin{figure}[t]
  \includegraphics[trim=0 0 0 0, clip,width=0.44\textwidth]{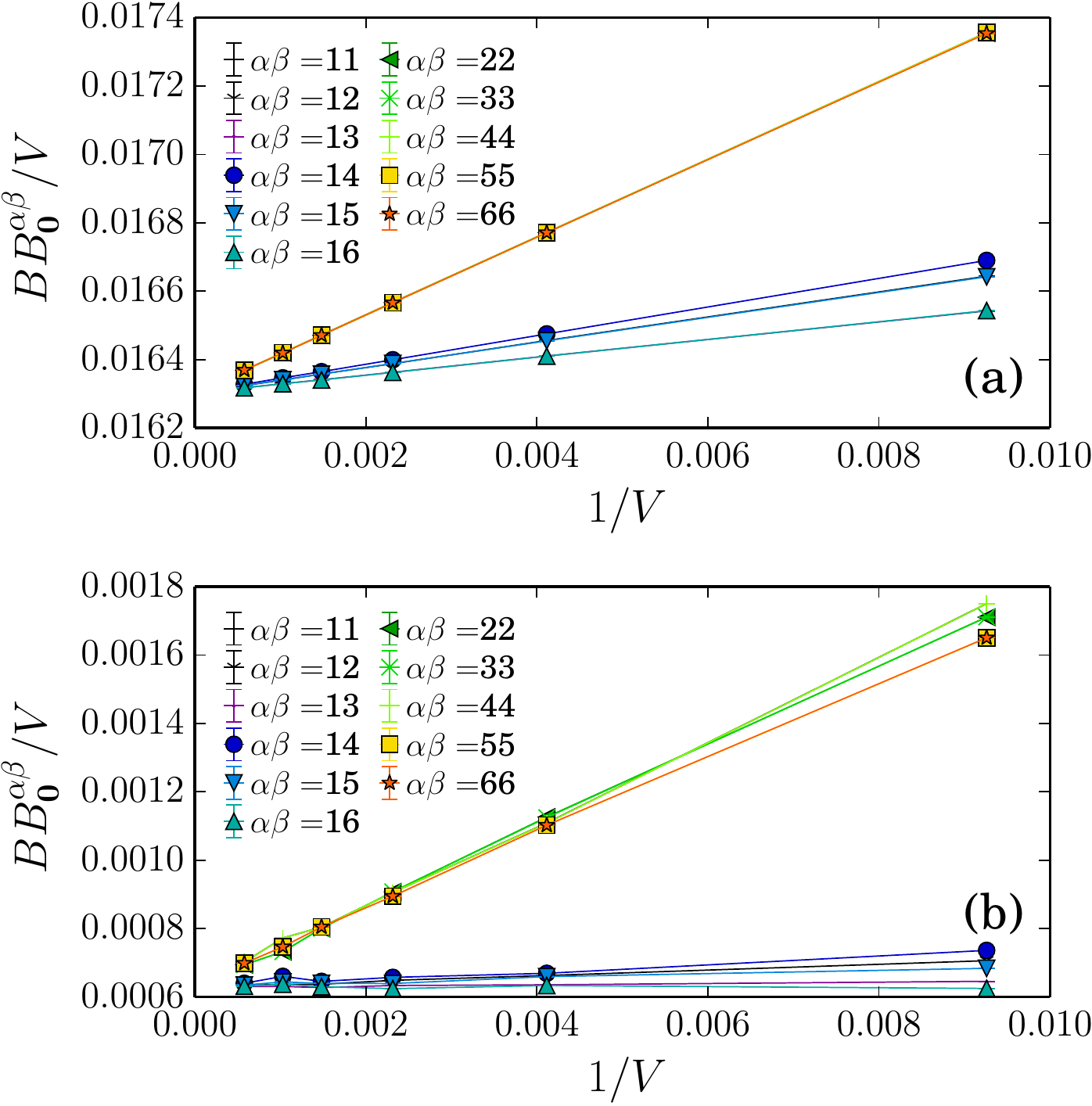}
  \caption{\label{fig:rotinv}The finite-size data for different $BB^{\alpha\beta}_{0}$ as a function of $1/V$ in the FM and in the lobe ((a) and (b), respectively).}
\end{figure}
In Fig.~\ref{fig:rotinv} we show the finite-size scaling of several matrix elements $B^{\alpha\beta}_{0}$  in the FM phase and in the lobes 
(Fig.~\ref{fig:rotinv}(a) and Fig.~\ref{fig:rotinv}(b), respectively). The finite-size scalings clearly suggest that the different $B^{\alpha\beta}_{0}$ converge to 
the same value as $V\to\infty$. Using a linear fit to the finite-sized data, we evaluate the extrapolated matrix elements $\lim \limits_{V\to\infty} BB^{\alpha\beta}_{0}/V$ 
in the FM phase and in the lobes, as shown in Fig.~\ref{fig:EXTrotinv}(a) and (b). These are consistent with a picture where there is no breaking of rotational invariance in both the FM and in the lobe phases.

\subsection*{Finite-size scaling inside the lobe for larger system sizes and  $T=J_z/16$.}
\begin{figure*}[b]
  \includegraphics[width=0.65\textwidth]{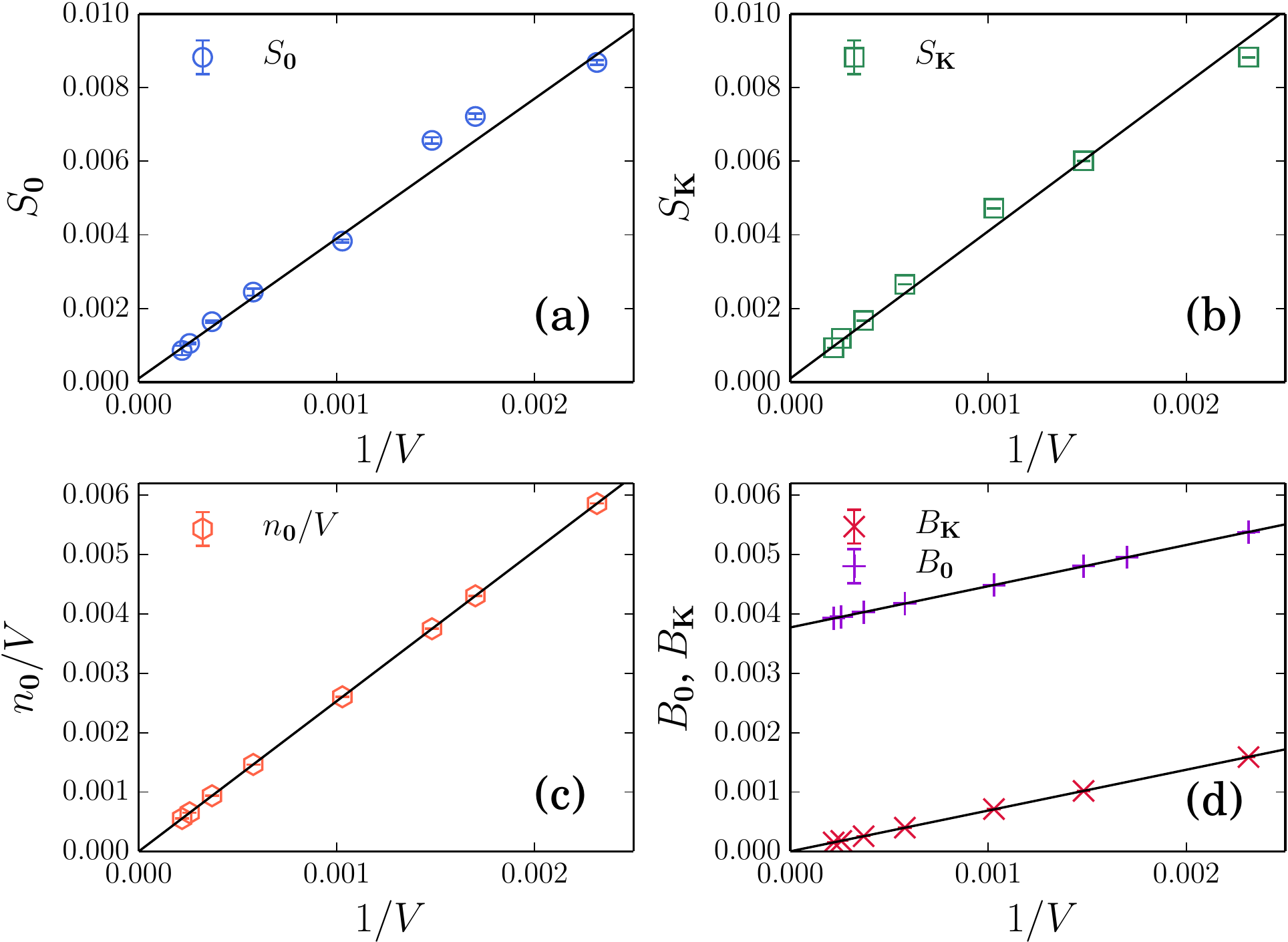}
  \caption{\label{fig:scal_beta16} Finite-size scaling of the structure factors $S_{\boldsymbol{0}}$ (a), $S_{\boldsymbol{K}}$ (b), $n_0/V$ (c),  $BB_{\boldsymbol{K}}$ and $BB_{\boldsymbol{0}}$ (d). The maximum system size in the figures corresponds to $V=39\times39\times3$. The temperature is set to  $T=J_z/16$, while $h/J_z=0.8333$ and $J_{\pm\pm}/J_z=0.495$.}
\end{figure*}

In Fig.~\ref{fig:scal_beta16} we re-examine the finite-size scaling of the most relevant candidate peaks in the structure factors presented in Fig.~\ref{fig:sfactors} 
using larger system sizes and higher temperature. The parameters of the simulations are $h/J_z=0.8333$, $J_{\pm\pm}/J_z=0.495$, and higher $T=J_z/16$. The largest cluster 
in the Fig.~\ref{fig:scal_beta16} corresponds to  $N_s=39\times39$, which happens to accomodate the two relevant reciprocal lattice vectors $\boldsymbol{0}$ and $\boldsymbol{K}$.
The continuous lines represent linear fits to the numerical data and the extrapolations to the limit of $V\to\infty$ are stable within the error bars 
upon removal of the smaller system sizes. These results obtained on larger clusters support the conclusion the phases in the lobes correspond to magnetically disordered phases.

\subsection*{Spin structure factors in the lobe at $T=J_z/96$. }
\begin{figure*}[b]
  \includegraphics[trim=0 0 0 0, clip,width=0.8\textwidth]{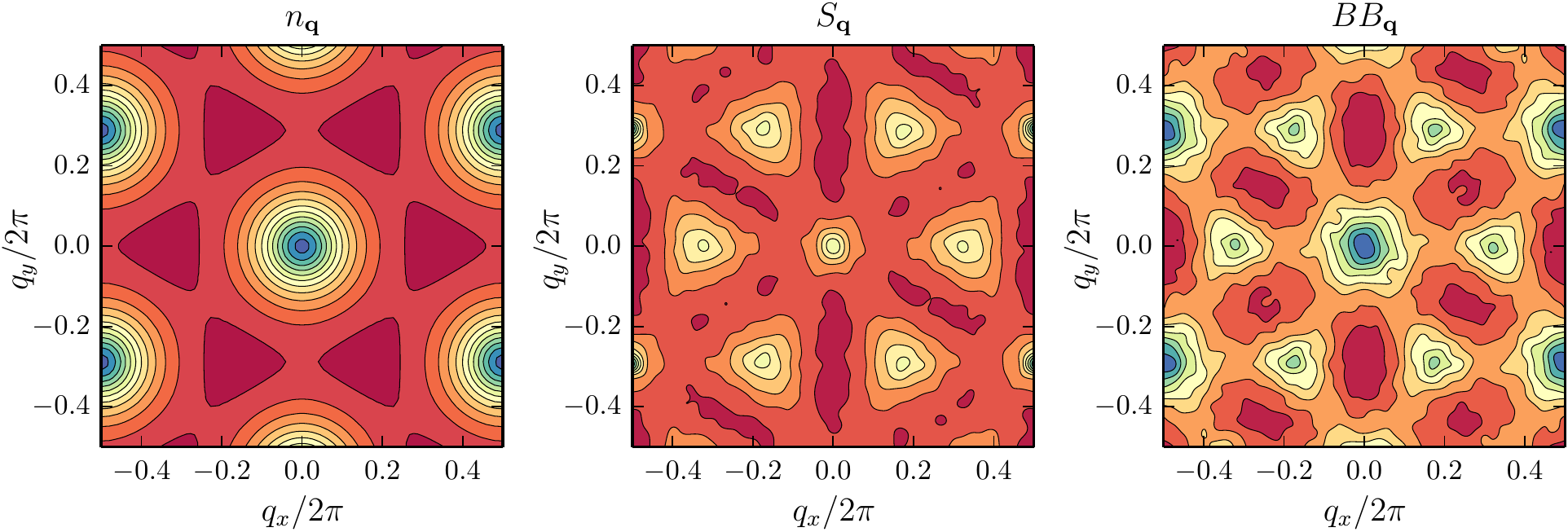}
  \caption{\label{fig:lowtstruc} Off-diagonal $n_{\boldsymbol{q}}$, diagonal  $S_{\boldsymbol{q}}$, and bond $BB_{\boldsymbol{q}}$ structure factors inside the lobe for a system with $N_s=24\times24$, $h/J_z=0.8333$, $J_{\pm\pm}/J_z=0.495$, and $T=J_z/96$.}
\end{figure*}

In figure~\ref{fig:lowtstruc} we present the structure factors inside the lobe for for a system with $h/J_z=0.8333$, $J_{\pm\pm}/J_z=0.495$, and $T=J_z/96$, on a cluster with
$N_s=24\times24$. The data confirm that at the lowest temperature reached in our simulations, the conclusions about the existence of a magnetically disordered phase in the lobes still holds.

\subsection*{Details of the FM phase and absence of a phase transition at zero field}

\begin{figure*}[b]
  \includegraphics[width=0.8\textwidth]{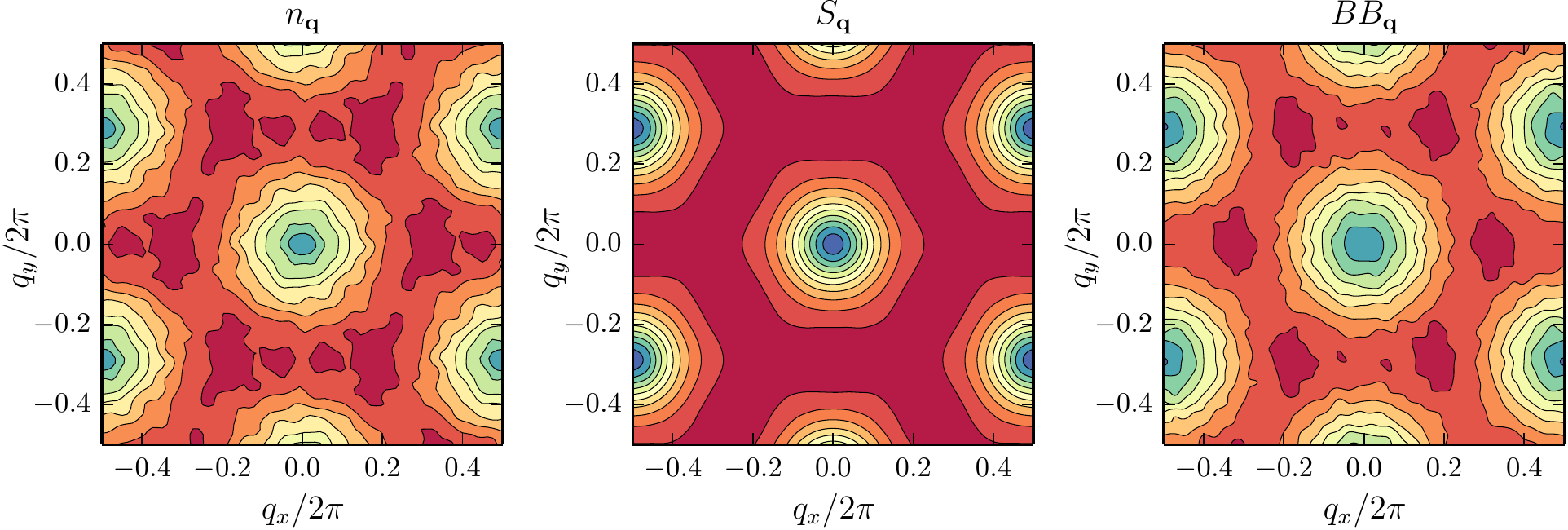}
  \caption{\label{fig:FMphase} Off-diagonal $n_{\boldsymbol{q}}$, diagonal  $S_{\boldsymbol{q}}$, and bond $BB_{\boldsymbol{q}}$ structure factors in the 
FM phase for a system with $N_s=24\times24$, $T=J_z/48$, and $h/J_z=0.8333$. The zero-momentum peaks of $n_{\boldsymbol{q}}$ and $BB_{\boldsymbol{q}}$ have been removed.}
\end{figure*}

\begin{figure}[t]
  \includegraphics[width=0.45\textwidth]{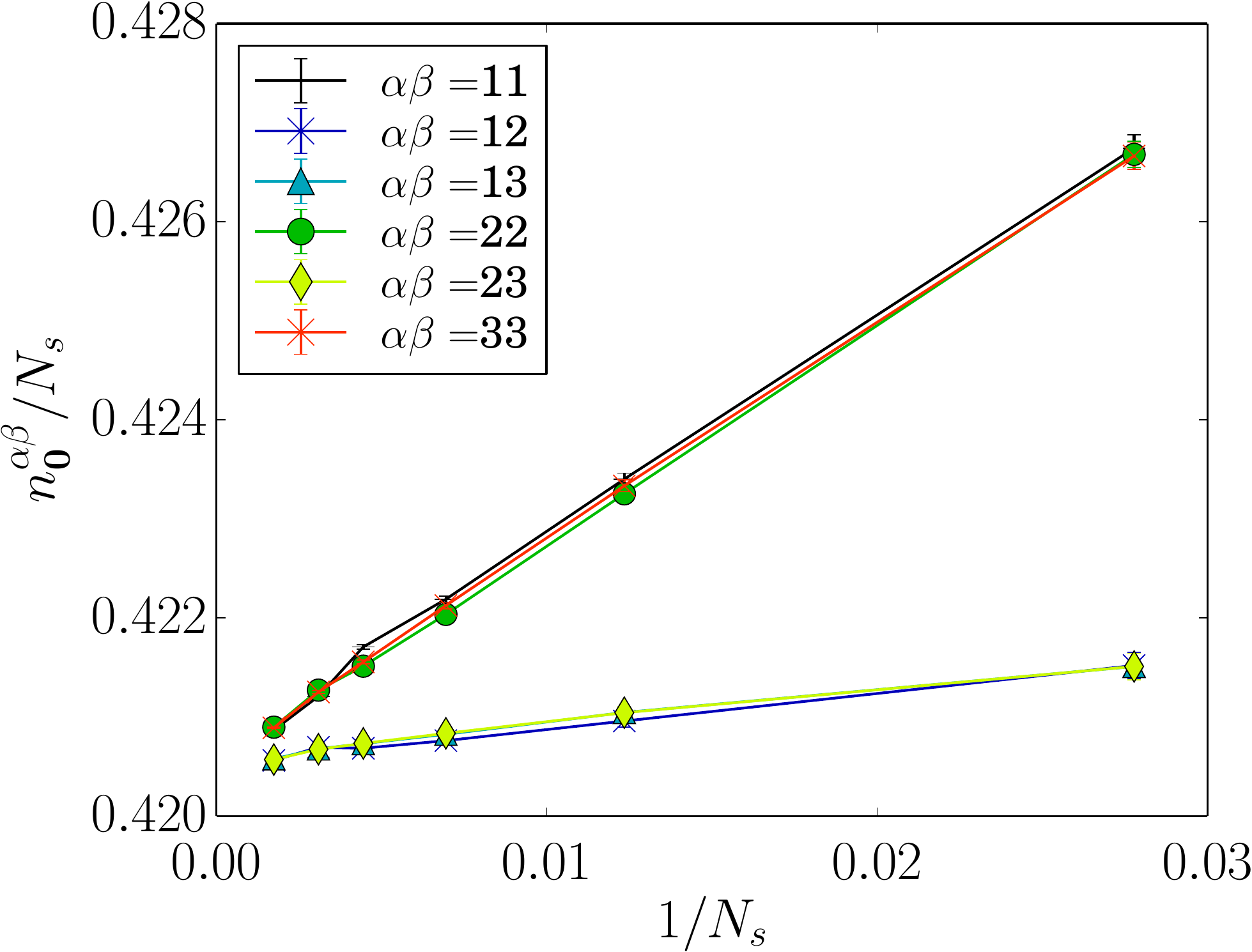}
  \caption{\label{fig:scaling_FMphase} The scaled sublattice zero-momentum occupation $n^{\alpha\beta}_{0}/N_s$ as a function of $1/N_s$. }
\end{figure}

In figure~\ref{fig:FMphase} we present results for spin correlation functions just outside the upper $1/3$-magnetization lobe. Both $n_{\boldsymbol{q}}$ and 
$BB_{\boldsymbol{q}}$ exhibit diverging peaks at zero momentum (which have been removed in the figures), while  the diagonal structure factor $S_{\boldsymbol{q}}$ 
does not. The fact that $f_0$ remains finite in outside the lobe together with the absence of diverging peaks in $S_{\boldsymbol{q}}$ means that the spins order in 
the $XY$ plane. 
\begin{figure}[t]
  \includegraphics[width=0.45\textwidth]{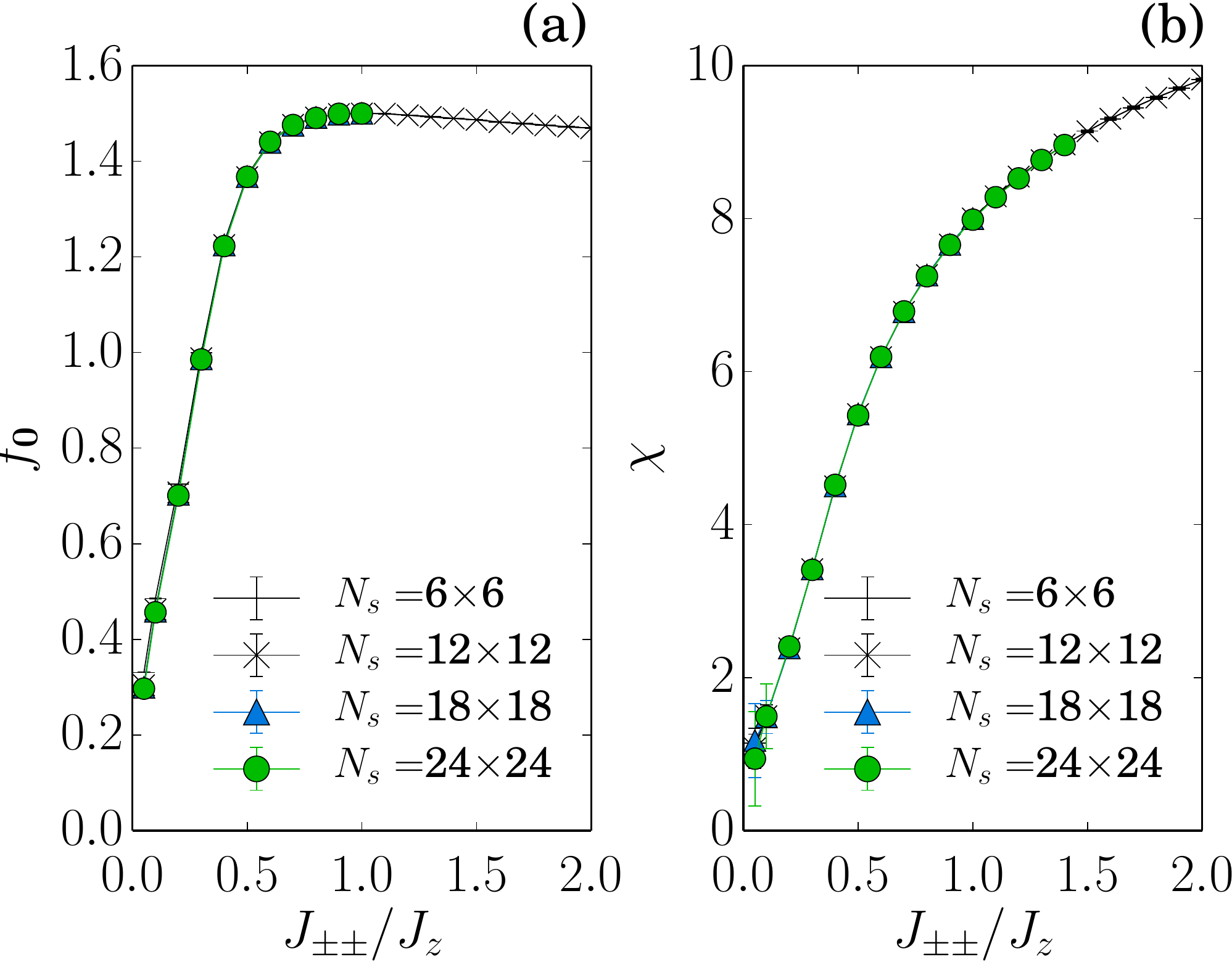}
  \caption{\label{fig:zero_field} The ``condensate fraction''  $f_0$ and the uniform magnetic susceptibility $\chi_z$, ((a) and (b),respectively), as a function of $J_{\pm\pm}/J_z$ and system size at zero magnetic field $h$ and $T=J_z/96$.} 
\end{figure}

To clarify the nature of this ordered phase, we analyze the sublattice zero-momentum occupation $n^{\alpha\beta}_{0}$ through finite-size 
scaling shown in figure~\ref{fig:scaling_FMphase}. In the limit of $V\to\infty$, we expect that in an ordered phase the matrix elements
$f^{\alpha\beta}_0\stackrel{\text{def}}{=}\lim\limits_{N_s\to\infty} n^{\alpha\beta}_{0}/N_s=\langle S_{\alpha}^{+}\rangle \langle S_{\beta}^{-}\rangle$, i.e., they
coincide with products of expectation values of the spin operators, which we extract from linear extrapolations of our data in figure~\ref{fig:scaling_FMphase}. The 
resulting matrix is
\begin{equation}
\left[\begin{array}{ccc}
0.42060 & 0.42053 & 0.42057 \\
0.42059 & 0.42053 & 0.42057 \\
0.42059 & 0.42053 & 0.42057 \\ 
\end{array}\right]\pm 10^{-5}\left[\begin{array}{ccc}
6 & 2 & 2 \\
6 & 2 & 2 \\
6 & 2 & 2 \\
\end{array}\right]
\end{equation} 
The fact that all matrix elements are real, positive, and have same magnitude means that the three spins in the unit cell are either aligned or antialigned along the $x$ 
direction (consistent with the two possibilities given the $Z_2$ symmetry of the Hamiltonian), thus the ground state is a ferromagnet. This is consistent with spin-wave 
theory calculations and with the expectation that in the limit of large 
$J_{\pm\pm} \gg J_z$, when the terms $-\left(S^{+}_{\boldsymbol{r}}S^{+}_{\boldsymbol{r'}}+S^{-}_{\boldsymbol{r}}S^{-}_{\boldsymbol{r'}}\right)=2\left(-S^{x}_{\boldsymbol{r}}S^{x}_{\boldsymbol{r'}}+S^{y}_{\boldsymbol{r}}S^{y}_{\boldsymbol{r'}} \right)$ dominate, the spins gain energy by aligning solely along the $x$ axis, where the spin interaction is unfrustrated.
Finally, the zero-field results for the order parameter $f_0$ and the uniform spin susceptibility are shown in Fig.~\ref{fig:zero_field}(a) and (b). The results clearly point 
towards the absence of a phase transition as $J_{\pm\pm}/J_z$ is decreased. The FM phase remains stable down to $J_{\pm\pm}/J_z=0$ where the extrapolated order parameter 
$f_0$ is about $11\%$ of its maximum value, which occurs exactly at the U(1) symmetric point $J_{\pm\pm}/J_z=1$. Similarly, the uniform susceptibility appears to be finite
in the limit $J_{\pm\pm}/J_z\to 0$. This scenario is similar to what happens in the spin-1/2 XXZ model on kagome lattice, where numerical evidence shows that a FM 
phase persists down to $J_{\pm}/J_z\to 0$,~\cite{Isakov2006} which was understood as the absence of vortex condensation in a duality analysis of the model.\cite{Isakov2006,Sengupta2006}

\end{document}